# Comments on indeterminism and undecidability

İnanç Şahin*

*Department of Physics, Faculty of Sciences,*

*Ankara University, Ankara, Turkey*

## Abstract

In a recent paper [1], it has been claimed that the outcomes of a quantum coin toss which is idealized as an infinite binary sequence is *1-random*. We also defend the correctness of this claim and assert that the outcomes of quantum measurements can be considered as an infinite *1-random* or *n-random* sequence. In this brief note we present our comments on this claim. We have mostly positive but also some negative comments on the arguments of the paper [1]. Furthermore, we speculate a logical-axiomatic study of nature which we believe can intrinsically provide quantum mechanical probabilities based on *1(n)-randomness*.

*inancsahin@ankara.edu.tr

In Ref.[1] author seems to prove that the outcomes of a quantum coin toss experiment, idealized as an infinite binary sequence is *1-random*. The author presents this result in Corollary 4.2 in his paper. He then concluded that deterministic hidden variable theories compatible with quantum mechanics should be completely ruled out. We think that the author's reasoning (his theorem 4.1 together with theorem B.3) provides a convincing evidence that quantum theory generates a *1-random* sequence of outcomes. However, there are some loopholes in his conclusion that determinism is ruled out. One issue that is often overlooked is the possibility that even a *1-random* sequence can be generated by an *unknown* algorithm. This *unknown* algorithm opens a door to determinism.[1]

Let me explain how this is possible. In 1961 Lucas [2] and in the early 1990s Penrose [3, 4] examined the problem of the equivalence of the human mind to a Turing machine. The summary of their conclusion was that the human mind surpasses the capabilities of the Turing machine. According to Gödel's incompleteness theorem or Turing's theorem for halting problem, all mathematical true formulas that a human mathematician can cognize, can neither be formalized in an axiomatic system nor described by an algorithm. However, as Penrose points out, there is an exception to this result: there can be an unknowable putative algorithmic procedure underlying mathematical understanding [4]. Gödel was also well aware of this fact. In his Gibbs Lecture in 1951, he pointed out that his incompleteness theorem do not preclude the possibility that there is a theorem-proving computer which is in fact equivalent to mathematical intuition [5]. However, the exact specification of the computer must be unknowable.[2] Consequently, according to this unknown algorithm the Gödel sentences which say of themselves in a metamathematical sense that they are not provable in Peano arithmetic can be deduced. Then, these Gödel sentences become decidable.[3] This unknown algorithm can also solve the halting problem. Hence, Chaitin's Ω which gives the probability that a completely random program will halt [6, 7]

$$\Omega = \sum_{p\ halts} 2^{-|p|} \qquad (1)$$

is no longer random according to this unknown algorithm. This unknown and unknowable algorithm makes *1-random* sequences *unknowable*. Consider for example, a Martin-Löf test

$$T = \{(i, \sigma) : i \in \mathbb{N},\ \sigma \in 2^{<\omega},\ c.e.\} \qquad (2)$$

which is a computably enumerable (c.e.) set of pairs of the form $(i,\sigma)$ of numbers $i$ and strings $\sigma \in 2^{<\omega}$ [7]. Define the sets

$$A_i = \bigcup_{(i,\sigma) \in T} N_\sigma \ ; \quad N_\sigma = \{h \in 2^\omega : \sigma \subset h\}. \tag{3}$$

Assume that $f \in 2^\omega$ passes all known tests, i.e. $f \notin \bigcap_{i \geq 0} A_i$ for all known $T$. Therefore, we may claim that f is *1-random*. However, we know that an unknowable algorithm may exist and $f$ can fail the test for c.e. sequence $\{A_i\}_{i \geq 0}$ of $\Sigma_1^0$ sets determined by this algorithm. Consequently, *1-random* property of any sequence remains unknown forever; the *genuine 1-randomness* cannot be grasped.[4]

The above analysis reveals the fact that there may be an unknown formal system $S^*$ encompassing all of our mathematical truths, and the sequences seem 1-random to us are not genuinely 1-random within this system. We can identify this unknown formal system with *nature*. We cannot apply Gödel's incompleteness theorem to $S^*$ and we cannot know that it is consistent or not, but its consistency is logically possible.[5] Such a formalization of nature can make physics deterministic.[6] Someone who defends an epistemic view might reject this argument by claiming that there is no reality beyond our knowledge. Therefore an unknowable algorithm cannot exist. Obviously our view is realistic; we accept the reality of the external world beyond us. On the other hand, the epistemic view does not necessarily rule out the existence of information, beyond our own knowledge. We think that the rejection of the existence of such an unknown algorithm, evokes solipsism.

Another interesting option is to think of nature as a system of an infinite and ever increasing number of axioms. Accordingly nature is an *open* formal system. If we represent the universe (or series of universes) as an infinite sequence of bits and somehow relate the order of the bits to *time*, then we might hope to obtain a *1-random* sequence.[7] To be precise, assume that nature is formalized in an infinite axiomatic system with infinite number of axioms. Moreover, the number of its vocabulary and the number of formation and transformation rules can be infinite. But for the sake of simplicity, assume that they are finite and static. Let us accept the reality of time and assume that new independent axioms are constantly being created.[8] For instance, consider a certain axiomatic system and write a Gödel sentence in that system. Then, that Gödel sentence or a proposition from which we can derive that Gödel sentence becomes an independent axiom and can be added to the system without causing any contradiction. Proceeding in this way, we expand the formal system by adding new independent axioms into the system. Such a process is infinite, it never ends. It generates an incomputable sequence of axioms. The creation of new independent axioms will expand the system in the same way that Gödel sentences are added as axioms.[9] How can the evolution of the universe be described in such a formal system? Let's consider the following scenario [10]: At every moment of time, the universe is represented by a long but finite sentence written according to the formation rules of the system. This sentence can be encoded as a sequence of binary digits. If it is a theorem of the formal system, then consider its shortest proof and the Gödel number associated with it. There is an axiom (or group of axioms) which we call evolution axiom. The evolution axiom processes the Gödel number of the shortest proof of the first sentence of the universe and gives a second Gödel number for another well-defined sentence.[11] Then, the universe evolves to the second sentence. On the other hand, if this first sentence is not a theorem of the system, then the behavior of the universe is undecidable.[12] In this case, the evolution freezes and waits for a new axiom to be created to prove the sentence. However we know that whether a given sentence is a theorem or not is generally undecidable within that system. The crux is to think of second, third, and higher order processes. Since we have accepted the reality of time, we can conceive processes that start from axioms and perform proofs of theorems, just like the operation of a working computer. These processes run over time and evolution freezes until the sentence of the universe is proven. Eventually, the evolution of the universe is incomputable and indeterministic.[13] As we know in quantum mechanics the evolution of a closed system (involved system + environment) is unitary and deterministic. On the contrary, in our model the universe evolves in an indeterministic way. However, the universe is not a closed system in our model. Could the incomputable evolution of the string that describes the universe be an inherited property that is also transferred to its substrings? If so then could the indeterministic and probabilistic structure of quantum mechanics have such an origin? We do not know the answers to these questions, but we think this big speculation should be examined.

This view can be challenged for many different reasons. First of all, the idea that nature is infinite is against our common sense. Considering nature as a formal system with infinite number of axioms possibly means infinite number of physical laws. Yet the world we see is finite. This seems to be an absurd sounding idea, it evokes the story attributed to B. Russell; *turtles all the way down* [8]. However, we think that most of the oppositions arise from psychological reasons. Moreover, the infinity problem could be solved by relational physics. For instance, consider two substrings of the universe say $O$ and $S$ which correspond to an observer and the system observed. Let the evolution of the substrings $O$ and $S$ over time be represented by the sequences $O_t$ and $S_t$ of the binary digits. If $O_t$ and $S_t$ are created by the same axioms or rules, then they follow the same pattern. However since the physics is relational, $O$ cannot derive this pattern from her observation. Accordingly, there may be an infinite number of axioms that we are not aware of; we can only recognize finitely many of them relationally. Although there are serious problems to solve, we think that the idea of considering nature as an open formal system can provide a basic understanding of the indeterministic and probabilistic structure of quantum mechanics.

Now let's continue our comments. We have demonstrated that 1-randomness of the outcomes of a quantum coin toss experiment, cannot be proved and cannot even be grasped intuitively. The phrase "quantum mechanics" has two different meanings. Its first meaning is the physics that characterize the microscopic world. By this we mean what is intrinsic in nature, not how we define it. Its second meaning is the formal theory of quantum mechanics which provides a language that allows us to understand nature. Since this formal system has finite number of axioms, it cannot intrinsically provide indeterminism and probabilities; these concepts are a matter of interpretation.[14] By denying the existence of an unknowable algorithm, one could argue that quantum mechanics gives us a sequence of outcomes in a way that does not depend on an algorithm at all, even if it is unknowable. Of course, such a claim is about the interpretation of probabilities. But what could such an interpretation be based on? Certainly, we cannot know the truth of such an interpretation.

Let's restrict ourselves to the case where there is no unknowable algorithm or formal system. Then, according to Theorem 5.1 of Ref.[1], deterministic theories that produce the same experimental results as quantum mechanics cannot exist. We consider this result to be correct. But in our opinion, there is another small point that we should discuss. We think that this result implicitly implies the infinity of time. The essential point is that a 1-random sequence must have infinite length. As far as I can understand, the outline of the author's reasoning is as follows: If we consider the infinite *idealized* sequence of outcomes generated by the formal system of quantum mechanics for a coin toss experiment (with an interpretation of the probabilities), it can be deduced that the outcomes form a 1-random sequence. On the other hand, deterministic theories cannot produce 1-random sequence of outcomes. Consequently, deterministic theories that produce the same experimental results as quantum mechanics cannot exist. Here we are actually comparing the quantum theory's prediction for an idealized infinite series of outcomes to that of a deterministic theory. One might think that the infinite sequence mentioned here is just an idealization and does not really have to exist. This thought would be correct if we were to question the existence of a deterministic theory that produces the same experimental results as the quantum theory (formal system + interpretation). However we should question the deterministic theory that produces the same sequence of outcomes as nature provides. Suppose for a moment that time is finite. Accordingly, the result of all quantum coin toss experiments obtained during the very long but finite life of the universe, creates a very long but finite sequence. In principle, there is always an algorithm that produces such a finite sequence. Therefore, we can always construct a deterministic theory with finite number of axioms.[15] Indeed, if we consider a sequence that repeats itself over a period at the order of the life time of the universe, such a sequence would appear random. It can be argued that for us, who are part of the universe, it is not possible to discover such an algorithm empirically, so such an algorithm can be considered to be the unknowable algorithm we mentioned earlier. However, this algorithm does not necessarily have to be the unknowable algorithm we mentioned earlier; it may not be equivalent to our mathematical intuition. Consequently, the empirical content of quantum mechanics can always be produced by a deterministic theory, unless we assume the infinity of time. Now assume that time is infinite and consider the infinite binary sequence of the outcomes of a quantum coin toss experiment. According to Chaitin's first information-theoretic incompleteness theorem [9], the formal system (deterministic theory) that generates a finite substring of this binary sequence should have an increasing complexity as the length of the substring increases.[16] Therefore, such an infinite binary sequence of the outcomes has infinite complexity and cannot be generated by a formal system with finite number of axioms.

As a result of efforts to explain the indeterministic and probabilistic nature of quantum mechanics, three different approaches can be adopted: (i) We may accept the existence of a finite formal system, possibly unknowable, which gives us a deterministic theory of nature. (ii) We may imagine an infinite formal system that can generate genuinely 1-random sequences, similar to what we discussed earlier. (iii) Finally we may assume that nature cannot be formalized in a (finite or infinite) formal system. In this case, probabilistic nature of quantum mechanics is a kind of mystical element that we cannot explain but only have to admit. Among these three approaches, approach (ii) can be seen as the least likely due to Occam's razor principle, as it provides an infinite complexity. However, we think that when we try to explain existence at the most basic level, it is not correct to rely on this principle.[17] Our experience that we live in a finite world cannot provide evidence that there can be no eternity in nature. As we mentioned earlier, relational physics can explain why we have finite experiences even though we are in an infinite nature. Moreover, we believe that approach (ii) can provide an explanation for the concept of absolute probability in quantum mechanics on the basis of undecidability. Essentially, undecidability and randomness are deeply related but different concepts. But then how can randomness arise as a result of undecidability? Of course, this requires some additional assumptions. Consider an infinite binary sequence, say $f \in 2^\omega$. Suppose the value of each bit of $f$ is undecidable; by this we mean that there is no effective procedure to determine whether the value of the bit is 0 or 1.[18] Let's pick a randomly chosen bit of $f$. Since there is no rule that dictates or forbids this element to be 0 or 1, our intuition tells that this chosen element will be 0 or 1 with a 50% -50% probability. It can be argued that the origin of our intuition is the totalitarian principle. Whatever its origin, it is clear that we indeed have such an intuition. Therefore, undecidability, at least intuitively, evokes the concept of probability. However, there is an important point we want to draw attention to. Undecidability for individual bits of the sequence does not guarantee the sequence to be 1-random. Consider for example the infinite string ...10101010... This infinite string is established according to a very simple rule: 1 is always followed by 0, and vice versa. Suppose the value of the first bit of this sequence is undecidable. This makes the value of each bit of the sequence undecidable. If we pick any bit from this infinite string, we would expect this bit to be 0 or 1 with probability 50%-50%. On the other hand, it is clear that aforementioned infinite string is not 1-random. Here, the essential point is that undecidability for individual bits does not mean that there are no rules between strings of bits. Indeed, consider length 2 substrings of this infinite sequence. These length 2 substrings obey the following rule: the sum of the digits of the substrings is always 1. Accordingly, in order to obtain a 1-random sequence, each substring of the sequence must be undecidable. A formal proof could be the following: Assume that $f \in 2^\omega$ is not 1-random. Then, there exist a c.e. test $T = \{(i,\sigma) : i \in \mathrm{N}, \sigma \in 2^{<\omega}\}$ such that $f \in \bigcap_{i \geq 0} A_i$. Since $f$ is an element of each set $A_i = \bigcup_{(i,\sigma) \in T} N_\sigma$, some substrings of $f$ form a c.e. collection. Hence, there should be an effective procedure that generates these substrings. Contrapositively, if for every collection of substrings of $f$ we do not have any effective procedure that generates the collection, then $f$ is 1-random. So does the approach (ii) provide such undecidability? We don't know the answer. But we have the hope that for some proper configuration of the formal system (for some proper choice of the formation-transformation rules, vocabulary etc.) this can be achieved.

Now suppose that time is infinite and quantum mechanics generates 1-random sequences regardless of their origin (we may accept approaches (ii) or (iii)). The origin of indeterminism in quantum mechanics is the state vector reduction. Quantum systems evolve unitary and their evolution is deterministic until a measurement is made. Many physicists consider state vector reduction to be a serious problem and as a solution they accept various interpretations and models that do not involve state vector reduction. For those who think in this way, what is essentially true is unitary deterministic evolution; state vector reduction is a kind of illusion. On the other hand, if we believe that the notion of 1-randomness is inherent in nature, then we can assume that quantum systems evolve in an indeterministic way by means of a 1-random sequence. The observed indeterminism and probabilities are consequences of this 1-random evolution. However, there is a subtle point here. Consider two quantum systems $S_1$ and $S_2$, and denote 1-random binary sequences representing their evolution by $(S_1)_t$ and $(S_2)_t$. If we assume that outcome of any measurement is a relational notion in quantum mechanics [11], then we need to deal with the concept of relative randomness between the sequences $(S_1)_t$ and $(S_2)_t$. Although $(S_1)_t$ and $(S_2)_t$ are both 1-random, they may not be 1-random relative to each other. Therefore, if we want to obtain a 1-random sequence of outcomes, the degree of the randomness of $(S_1)_t$ and $(S_2)_t$ should be different. For example, if $(S_1)_t$ is *n-random* but $(S_2)_t$ is *(n+1)-random*, then $(S_1)_t$ is *1-random* relative to $(S_2)_t$ and vice versa. It is possible to explain such relative random behavior within the approach (ii). For this it should be accepted that the proofs of the sentences for systems $S_1$ and $S_2$ require the use of different axioms. We know that the universe is made up of certain kinds of particles. For this reason, the sentence belonging to the universe must be in the form of a collection of small sentences that are very similar to each other. Hence, atoms or groups of atoms in $S_1$ and $S_2$ should be written in similar sentences, and thus we can expect them to be proved by the same axioms. Indeed, it is very likely that the sequences $(S_1)_t$ and $(S_2)_t$ have a behavior determined by the same axioms. When this happens the variation in their code will be the same.[19] But the same change that occurs in every subsentence of the universe is not observable. An enormous amount of time may passed in this way, but observers cannot be aware that time has passed. A measurement identify a moment of time, but time will always remain undetermined; we cannot know our position on the timeline. The slight difference between the sentences of $S_1$ and $S_2$ sometimes requires different axioms to prove them. When this happens, relative random behavior occurs between $S_1$ and $S_2$.

Finally, we would like to mention an issue that weakens the argument of approach (i). If the locality and experimenters' free will are assumed, it can be shown that Bell's theorem rules out any algorithm that predicts experimental results even if algorithm is unknowable.[20] This simple result can be easily demonstrated. For this purpose, consider an EPR-type experiment with two spin-1/2 particles in the singlet state. Let A and B be spacelike separated observers with Stern-Gerlach apparatuses. $\bar{n}_A$ and $\bar{n}_B$ represent spin projection axes at A and B. Let the spins become entangled at the midpoint O of A and B. Then they are separated into points A and B in a process that preserves angular momentum. Suppose there is an algorithm that determines the measurement results at A and B when the particles become entangled at point O. Now let's choose a hidden variable which encodes this algorithm and assume that there are hidden universal Turing machines at points A and B. These Turing machines read the hidden variable when the particles arrive and output the measurement results. These outputs of Turing machines must obey Bell inequalities. This is obvious because such an algorithm and the hidden variable carrying it cannot transmit the spooky action at a distance. Since quantum mechanics can violate Bell's inequalities, the sequence of outcomes it produces cannot be generated even by an unknowable algorithm. For this reason, the unknowable algorithm cannot save determinism under the assumptions of locality and free will.

---

[1] We should note that we use the term determinism to mean superdeterminism or nonlocal determinism. Obviously, Bells theorem rules out local determinism if the experimenters' free will is assumed.

[2] Both Penrose and Gödel found this possibility highly unlikely, although it cannot be ruled out logically. However, we are only interested in its logical possibility and we do not use it in the context of mental computabilism.

[3] One might ask, what will happen if we try to write a Gödel sentence in the extended axiomatic system that includes our mathematical understanding. Since this axiomatic system is unknown we cannot write a Gödel sentence in that system and even though we could we cannot see its truth.

[4] This is a stronger statement than saying 1-randomness is unprovable. Because 1-randomness cannot even be grasped intuitively.

[5] It may not make sense to question the completeness of such a system. When we say a formal system is complete, we imply that all *true* formulas obtained by formation rules can be deduced from axioms by transformation rules. But since $S^*$ contains all of our mathematical truths as theorems, it is not possible to write a true formula which is not a theorem. Hence, $S^*$ is complete by definition. On the other hand, one may ask a weakened version of completeness: Can all formulas (sentences written with the correct grammar) be deduced from axioms as theorems? Since there may be some unknown formation and transformation rules, the answer remains unknown in general.

[6] My personal opinion is that such a determinism is not a correct description of nature, for various reasons, which I will discuss later. But determinism always remains a logical possibility, it cannot be completely eliminated. On the other hand, we can accumulate evidences supporting that nature is not deterministic. I consider the argument of Ref.[1] as such an evidence.

[7] The underlying philosophy is that epistemological study of the universe is possible and the universe and everything in it can be reduced to a sequence of 0's and 1's. However, the universe (or series of universes) is open in time and we have an infinite sequence of binaries. On the other hand, this should not be misunderstood; we do not advocate an epistemic approach. Furthermore, we do not claim that this binary structure is the essence of nature. But we assume that the universe is discrete, hence it can be *represented* by binaries.

[8] The direction in which the number of axioms increases determines the arrow of time.

[9] It can be asked, according to which algorithm these Gödel sentences are written in the axiomatic system. Obviously, we assume that there is no such algorithm. We give such an example just to make it easier to imagine. The actual process is that new independent axioms are created without any rules. Accordingly, as time progresses, some undecidable sentences of the system become decidable.

[10] We do not claim that this scenario describes the reality. We are just trying to imagine a scenario of what it might be like.

[11] Here, we should note that the evolution axiom acts depending on the Gödel numbers of the proof of theorems. So transformation rules are not used in this inference.

[12] One might ask whether the creation of axioms is based on some unknown algorithm. If this is the case, then this means axioms are not genuinely independent. Thus, the whole formal system can be derived from finite number of axioms. This gives a finite and static system, which contradicts with our initial assumption. Therefore, there should not be a rule or algorithm that dictates the creation of axioms.

[13] Determinism and computability are related but different concepts. Incomputability does not require indeterminism in general. On the other hand, we assume that the universe is discrete and hence, it is represented by a sequence of bits. Therefore any function of hidden variables $f : \Lambda \to R$ can be considered as a discrete function. Consequently, if we don't have any effective procedure which maps the elements of the discrete set $\Lambda$ to the elements of the discrete set $R$, then we don't also have a discrete function $f$.

[14] In contrast, the infinite dynamical axiomatic system we discussed in the previous pages has the potential to provide intrinsically indeterminism and probabilities.

[15] According to Chaitin's first information-theoretic incompleteness theorem [9] such a deterministic theory must have an extraordinary complexity. This probably means that this theory has an enormous number of axioms. But ultimately the number of axioms must be finite.

[16] In fact, it is not entirely clear that such a result is a necessary consequence of Chaitin's theorem. As it was shown in Ref.[10] the value of a real limiting constant does not have any connection to the complexity of axioms. On the other hand, since a genuinely 1-random sequence cannot be produced by a finite formal system, it seems correct to me that the deterministic theory has increasing complexity as the complexity of the string it produces increases.

[17] Occam's razor or Solomonoff's principle can be considered as a valid principle during the evolution of the binary string of the universe. But we do not consider it as a principle that limits the number of laws of nature.

[18] In fact, for the elements of the sequence with a finite index, an algorithm that derives these elements can always be constructed. When we say that each bit of $f$ is undecidable, we mean that there is no effective procedure that determines a bit with an arbitrarily large index. But for a moment, let's ignore this detail.

[19] This may not always be true. Slight differences between the sentences of $S_1$ and $S_2$ can cause the same axioms to evolve these sentences differently. But such a difference is computable and does not lead to randomness; in this way, perhaps some exact laws of physics can emerge.

[20] Actually, this result is not surprising, since indeterminism requires incomputability.